\documentclass{blois}

\def\be{\begin{equation}}
\def\ee{\end{equation}}
\def\bea{\begin{eqnarray}}
\def\eea{\end{eqnarray}}

\newcommand{\Photo}{}

\begin{document}
\vspace*{4cm}
\title{First Constraint on the Diffuse Supernova Neutrino Background through the CE$\nu$NS process from the LZ experiment}

\author{ Qing Xia (on behalf of the LZ Collaboration\footnote{https://lz.lbl.gov}) }

\address{Lawrence Berkeley National Laboratory (LBNL), Berkeley, CA 94720-8099, USA
\\\normalfont{Email:qingxia@lbl.gov}}

\maketitle\abstracts{
We report the limits on the diffuse supernova neutrino background (DSNB) flux and the fundamental DSNB parameters measured from the first science run of the LUX-ZEPLIN (LZ) experiment, a dual-phase xenon detector located at the Sanford Underground Research Facility in Lead, South Dakota, USA. This is the first time the DSNB limit is measured through the process of the coherent elastic neutrino-nucleus scattering (CE$\nu$NS). Using an exposure of 60~live days and a fiducial mass of 5.5~t, the upper limit on the DSNB $\nu_x$ (each of $\nu_\mu$, $\nu_\tau$, $\bar\nu_\mu$, $\bar\nu_\tau$) flux is $686-826$~cm$^{-2}$s$^{-1}$ at the 90\% confidence level for neutrino energies E$>$19.3~MeV, assuming the flux for each $\nu_x$ flavor is the same. The interval accounts for the uncertainty in existing DSNB models. The present result is comparable to the existing best limit and further improvements are expected after collecting data from an estimated 1,000-day exposure in the future.}

\section{Introduction}
The diffuse supernova neutrino background (DSNB) is theorized to be a nearly isotropic flux of neutrinos cumulatively originating from all past core-collapse supernovae. Each supernova core collapse is predicted to produce 10$^{58}$ neutrinos on average\cite{Suliga2020}. The detection of the DSNB is the only feasible method to probe the average neutrino emission per core-collapse event, as there has not been a nearby core-collapse event since SN 1987A, which occurred 37 years ago. Furthermore, even if another such event were to occur in the near future, there is no guarantee it would represent a typical supernova~\cite{Suliga:2021hek}. The understanding of core collapse depends on probing the DSNB in all flavors. The current
best upper limit on the $\bar\nu_{e}$ DSNB flux is 2.7~cm$^{-2}$s$^{-1}$, measured by Super-Kamiokande (SK), and it is close to a discovery~\cite{Super-Kamiokande:2021jaq,Beacom:2003nk}. The upper limit on the $\nu_e$ flux is 19~cm$^{-2}$s$^{-1}$, set by the Sudbury Neutrino Observatory (SNO)~\cite{SNO:2020gqd}, and will be improved by the future Deep Underground Neutrino Experiment (DUNE)~\cite{Zhu:2018rwc,Moller:2018kpn}. However, the best upper limit on the DSNB flux of $\nu_x$, i.e., each of $\nu_\mu$, $\nu_\tau$, $\bar\nu_\mu$, $\bar\nu_\tau$, measured using neutral current neutrino-electron scattering from SK is very weak~\cite{Lunardini:2008xd}, at the level of approximately 10$^3$~cm$^{-2}$s$^{-1}$, due to this detection channel's small interaction cross section.

New ideas have been proposed to enhance the DSNB sensitivity to all flavors~\cite{Suliga:2021hek}. One of these ideas involves probing the DSNB using direct-detection dark-matter experiments through the CE$\nu$NS process, where a neutrino interacts with a nucleus via the exchange of a $Z$ boson. The COHERENT experiment has made the first detection of the neutrino CE$\nu$NS process, which is sensitive to all flavors of neutrinos and antineutrinos~\cite{COHERENT:2017ipa}. The CE$\nu$NS process dominates in medium-A nuclei for neutrino energies less than about 50~MeV\cite{Scholberg:2018vwg}, including the DSNB which consists of $\mathcal{O}$(MeV) thermal neutrinos. Therefore, CE$\nu$NS is the most promising channel for xenon-based detectors to probe the DSNB. We report here the first upper limit on the DSNB flux measured through the CE$\nu$NS channel from the LUX-ZEPLIN (LZ) experiment.

\section{The DSNB Signal Model}
Depending on the progenitor mass, a core-collapse supernova will either lead to the formation of a neutron star (NS) or a black hole (BH). The neutrino flux from the BH formation is expected to be larger than that from NS formation, and the spectrum is expected to be harder because the collapsed cores of BH forming supernovae have higher mass and can generate more accretion onto them. The DSNB flux for a single neutrino flavor, in units of [cm$^{-2}$s$^{-1}$MeV$^{-1}$], can be written as~\cite{Suliga:2021hek}

\begin{equation}
    \Phi(E_\nu)=\frac{c}{H_0}\int_{8M_\odot}^{125M_\odot} dM \int_{0}^{z_\mathrm{max}}dz\frac{R_\mathrm{SN}(z,M)}{\sqrt{\Omega_\mathrm{M}(1+z)^3+\Omega_{\Lambda}}}\times[f_\mathrm{NS}F_\mathrm{NS}(E'_\nu,M)+f_\mathrm{BH}F_\mathrm{BH}(E'_\nu,M)],
\end{equation}

\noindent where supernova progenitor masses ($M$) between $8-125$ solar masses that result in either NS or BH formation are considered. The neutrino emission spectrum $F(E'_\nu,M)$ per core collapse is predicted by 1D hydrodynamical simulations with Boltzmann neutrino transport from the Garching group~\cite{Garching}. $E'_\nu$ is the neutrino energy at emission, and is related to the neutrino energy at detection, $E_{\nu}$, by $E'_{\nu} = E_{\nu}(1+z)$, where $z$ is the red-shift. The supernova rate density for a given mass is $R_{SN}(z,M)$. Standard values are used for the speed of light $c$, Hubble constant $H_0$, dark matter fraction $\Omega_M$, and dark energy fraction $\Omega_\Lambda$~\cite{10.1093/ptep/ptaa104}. 

\begin{figure}
\centering
\includegraphics[width=0.48\textwidth]{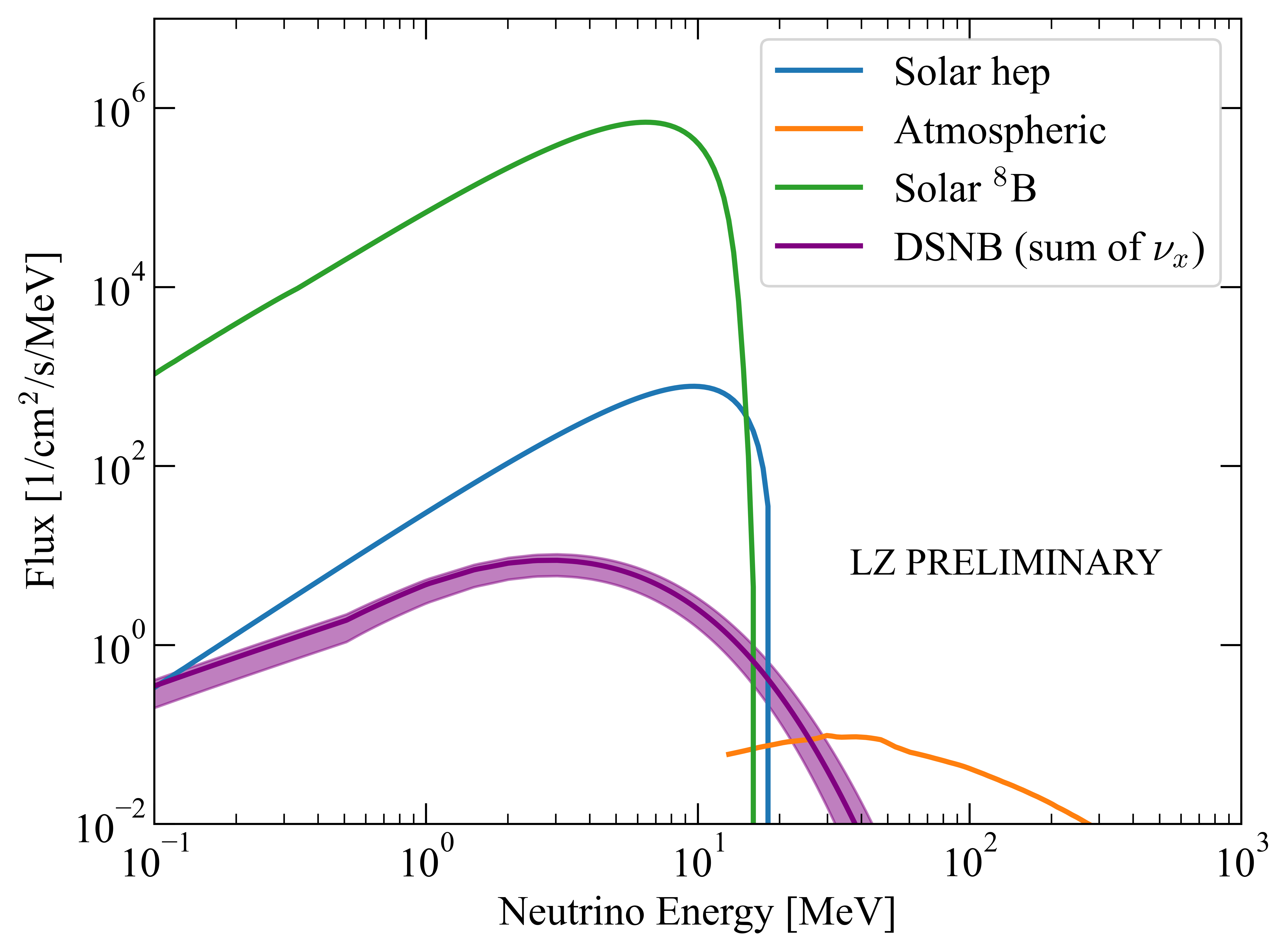}
\includegraphics[width=.48\textwidth]{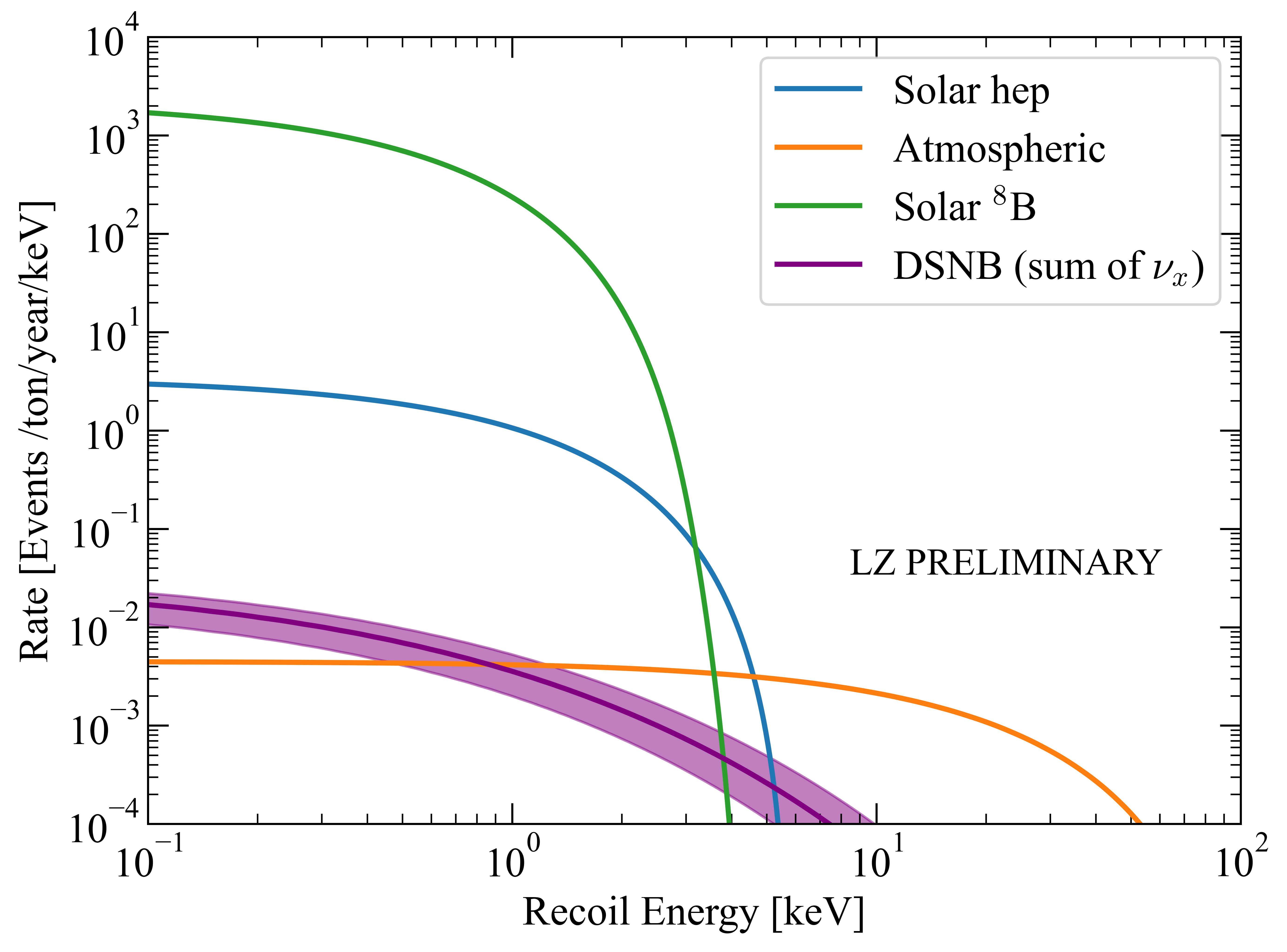}
\caption[]{Left: Predicted DSNB signal spectrum (sum of $\nu_x$) and irreducible neutrino backgrounds. The solid purple line represents the fiducial DSNB model, and the purple band indicates the uncertainty set by the minimal and maximal DSNB models in Ref.~\cite{Suliga:2021hek}. Right: Predicted nuclear recoil spectrum in xenon for the DSNB (sum of $\nu_x$) and irreducible backgrounds corresponding to the left plot.}
\label{fig:dsnbflux}
\end{figure}

Following Ref.~\cite{Suliga:2021hek}, the fiducial DSNB model used in this work has $R_{SN}(z=0)~=~ 1.25\times10^{-4}$~Mpc$^{-3}$yr$^{-1}$ and $f_{BH}\simeq$~20\%. For the minimal DSNB model, the pre-factor of $R_{SN}$ is changed to 0.75 and $f_{BH}$ is chosen to be $\simeq$~10\% (fast-forming BH model). For the maximal DSNB model, the pre-factor for $R_{SN}$ becomes 1.75 and $f_{BH}\simeq$~40\% (slow-forming BH model). The adopted model uncertainty is based on a conservative evaluation of the observed $R_{SN}(z=0)$ and $f_{BH}$ from astronomical surveys. The total flux of DSNB $\nu_x$ (sum of $\nu_\mu$, $\nu_\tau$, $\bar\nu_\mu$, $\bar\nu_\tau$), as well as solar and atmospheric neutrinos and their nuclear recoil spectra in xenon are shown in Fig.~\ref{fig:dsnbflux}. 

\section{Data Analysis}\label{sec:ana}
\begin{figure}
\begin{minipage}{0.5\linewidth}
\centerline{\includegraphics[width=1\linewidth]{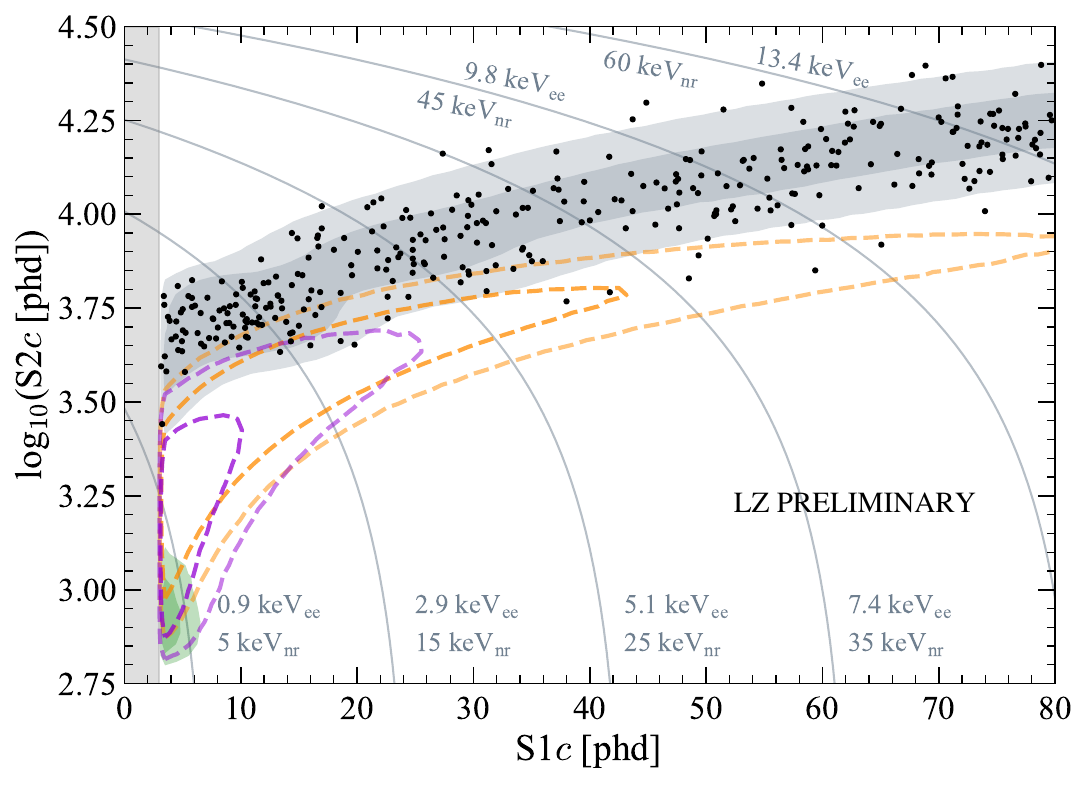}}
\end{minipage}
\hfill
\begin{minipage}{0.5\linewidth}
\resizebox{\textwidth}{!}{
\begin{tabular}{p{3.5cm}p{2.9cm}p{1.9cm}}
    \hline
    \hline
  Source & Expected Events & Fit Result\\
    \hline
    $\beta$ decays + Det. ER & 215 $\pm$ 36& 222 $\pm$ 16\\
    $\nu$ ER& 27.1 $\pm$ 1.6 & 27.1 $\pm$ 1.6\\
    $^{127}$Xe & 9.2 $\pm$ 0.8 & 9.3 $\pm$ 0.8\\ 
    $^{124}$Xe& 5.0 $\pm$ 1.4 & 5.2 $\pm$ 1.4 \\
    $^{136}$Xe &	15.1 $\pm$ 2.4\phantom{0} & 15.1 $\pm$ 2.4\phantom{0} \\
    Atmospheric + $^8$B \newline+ $hep$ CE$\nu$NS & 0.18 $\pm$ 0.02 & 0.18 $\pm$ 0.02 \\
    Accidentals	& 1.2 $\pm$ 0.3 & 1.2 $\pm$ 0.3 \\
    \hline
    \rule{0pt}{1.0ex} Subtotal & 273 $\pm$ 36\phantom{0} 
    & 280 $\pm$ 16\phantom{0} \\
    \hline 
    $^{37}$Ar & [0, 288] & $52.6^{+9.6}_{-8.9}$\phantom{0} \\[0.25ex]
    Detector neutrons &	$0.0^{+0.2 }$ & $0.0^{+0.2}$ \\[0.25ex]
    DSNB $\nu_x$ all flavors & -- & $0.0^{+0.5}$ \\[0.25ex]
    \hline
    Total & -- & 333 $\pm$ 17\phantom{0} \\
    \hline
    \hline
    \end{tabular}}
\end{minipage}
\caption[]{Left: 60 live days of data from LZ's first science run (black dots) after all cuts in log$_{10}$S2$c$-S1$c$ space. Contours enclose the 1$\sigma$ and 2$\sigma$ regions of the following models: the best-fit background model (shaded gray regions), the atmospheric $\nu$ CE$\nu$NS component (dashed orange lines), the fiducial DSNB signal (dashed purple lines), and $^8$B solar neutrinos (shaded green regions). Thin gray lines indicate contours of constant energies. Right: Number of background events from various components. The second column is the predicted number of events with uncertainties as described in Ref.~\cite{LZ:2022lsv}. The uncertainties are used as constraint terms in a combined fit of the background model plus the DSNB signal (fiducial model) to the selected data. The fit result is shown in the third column. $^{37}$Ar and detector neutrons have non-Gaussian prior constraints and are summed up separately.\label{fig:2dfit}}
\end{figure}

Data presented in this paper are from the first science run of LZ, which were collected from 23 Dec 2021 to 11 May 2022 under stable detector conditions. The data selection is identical to that used in the WIMP search analysis~\cite{LZ:2022lsv}, which results in a total livetime of 60$\pm$1~days and a fiducial volume LXe mass of 5.5$\pm$0.2~t. The background model used for the DSNB search is the same as for the WIMP search, with the addition of two neutrino CE$\nu$NS background components from atmospheric and solar $hep$ neutrinos. The expected number of events from these neutrino backgrounds, as well as all the other backgrounds in LZ's first science run are listed on the right of Fig.~\ref{fig:2dfit}.

Statistical inference of the DSNB $\nu_x$ flux is performed with an unbinned profile likelihood statistic in the log$_{10}$S2$c - $S1$c$ 2D parameter space with a two-sided construction of the 90\% confidence bounds~\cite{Baxter:2021pqo}, as shown in Fig.~\ref{fig:2dfit} (left). The black dots in the figure represent the 335 events passing all selections. The contours indicating the fiducial DSNB signal and the best-fit background model are also shown.

\section{Results and Prospects}
\begin{figure}[!ht]
\begin{minipage}{0.5\linewidth}
\includegraphics[width=1.0\linewidth]{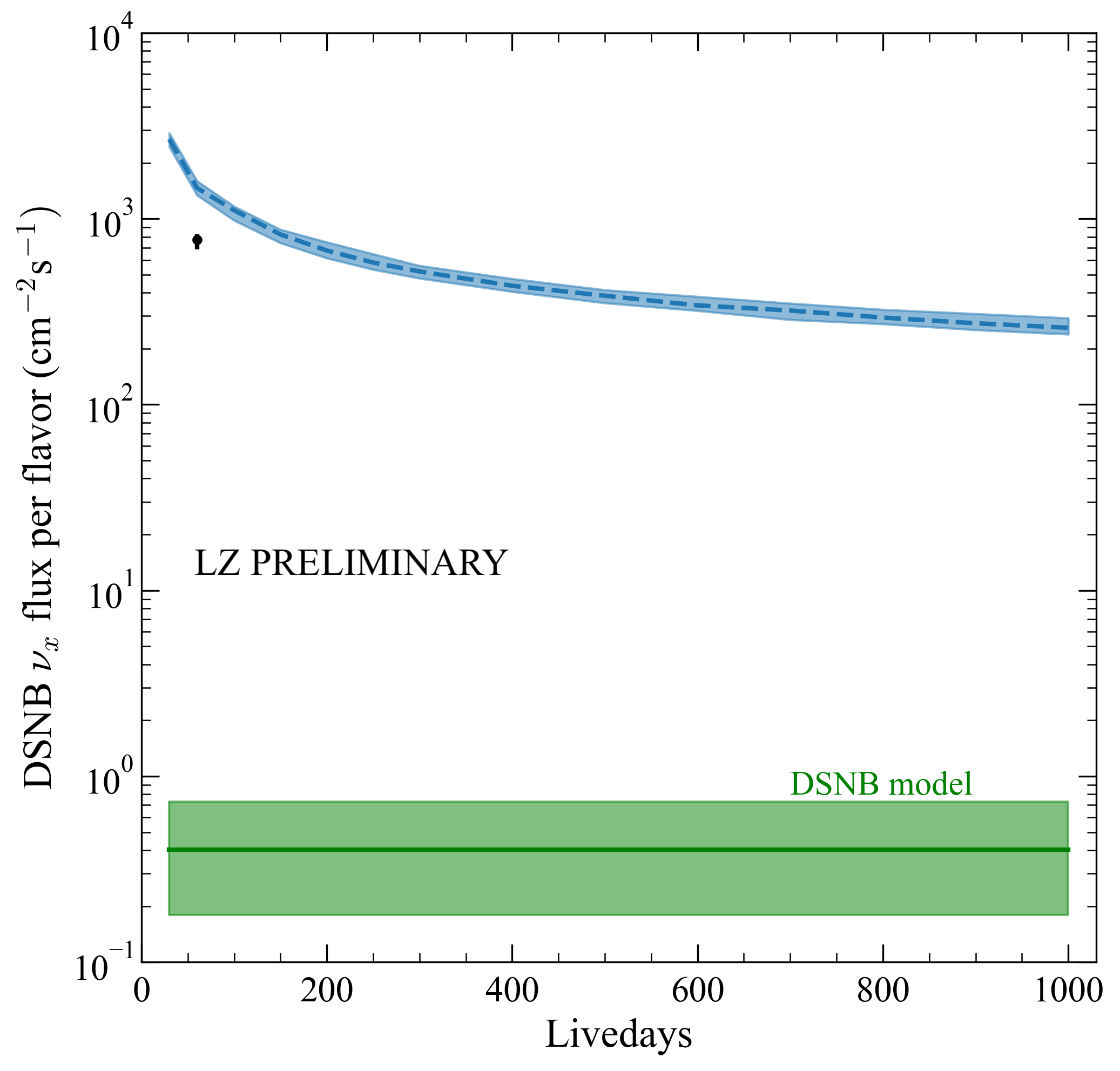}
\end{minipage}
\begin{minipage}{0.5\linewidth}
\includegraphics[width=1.0\linewidth]{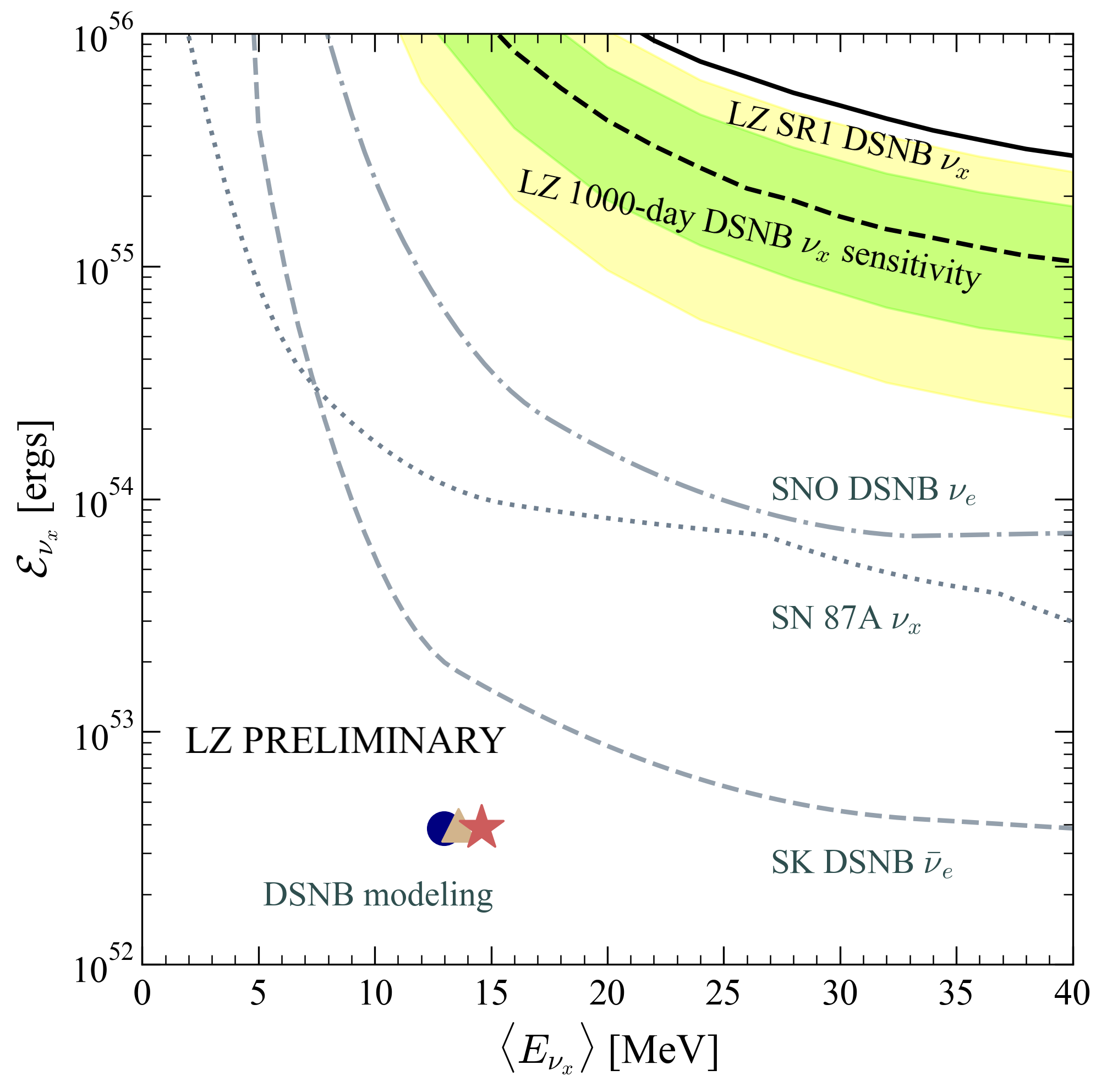}
\end{minipage}
\caption[]{Left: The 90\% C.L. upper limit on the DSNB $\nu_x$ flux per flavor (black) for neutrino energy $>$19.3~MeV, using 60 live days of data from LZ. The dot is the upper limit on the fiducial DSNB model, with its uncertainty set by the minimal and maximal models from Ref.~\cite{Suliga:2021hek}. The dashed blue curve is the projected median sensitivity to the DSNB $\nu_x$ flux (fiducial model) with LZ as a function of the detector live time. The horizontal solid green line is the predicted $\nu_x$ flux from DSNB modelling (fiducial model). Both bandwidths in blue and green are also set by the minimal and maximal model. Right: Limits on the fundamental DSNB $\nu_x$ emission parameters: the total emitted energy per $\nu_x$ flavor and average neutrino energy.
  The solid black curve is the limit from LZ's 60 live days of data in the first science run. The dotted line is the median of the 1,000-liveday sensitivity projection. The green and yellow bands are the 1$\sigma$ and 2$\sigma$ sensitivity bands. Also shown are the present SK limit on DSNB $\bar{\nu}_e$ (dashed gray)~\cite{Super-Kamiokande:2021jaq}, the SNO limit on DSNB $\nu_e$ (dashdotted gray)~\cite{SNO:2020gqd}, and the SN 1987A limit on $\nu_x$ (dotted gray)~\cite{Suliga:2021hek}, as well as three points that indicate the average emission per supernova collapse in the fiducial, minimal, and maximal DSNB models.
\label{fig:dsnbresult}}
\end{figure}
The results from the statistical analysis described in Section~\ref{sec:ana} show the data collected is consistent with the background-only hypothesis, as the best-fit number of DSNB signals is zero. The 90\% confidence level (C.L.) upper limit on the DSNB $\nu_x$ flux per flavor for neutrino energies $>$19.3~MeV is shown in Fig.~\ref{fig:dsnbresult} (left). The energy window is chosen to match that from Ref.~\cite{Suliga:2021hek,Lunardini:2008xd}. Because stringent constraints have already been placed on $\nu_e$ and $\bar{\nu}_e$ by SNO and SK, the limit on the DSNB $\nu_x$ flux is derived by neglecting the presence of the $\nu_e$ and $\bar{\nu}_e$ components of the DSNB and considering four $\nu_x$ flavors only, while assuming each of the four flavor fluxes is the same (following the conventions in Ref.~\cite{Suliga:2021hek}). This results in a 90\% C.L. upper limit on the DSNB $\nu_x$ flux of 773
~cm$^{-2}$s$^{-1}$ per flavor, assuming the fiducial DSNB model, and $686-826$~cm$^{-2}$s$^{-1}$ per flavor after accounting for the model uncertainty between the minimal and maximal models~\cite{Suliga:2021hek}. Due to downward fluctuations in the number of background events, the limits above are power constrained to the -1$\sigma$ range of the projected sensitivity following Ref.~\cite{cowan2011powerconstrainedlimits}. The projected sensitivity to the DSNB $\nu_x$ flux as a function of the detector live time for a fiducial mass of 5.5 tonnes assuming similar background levels as LZ's first science run is also shown in Fig.~\ref{fig:dsnbresult} (left). 

Figure~\ref{fig:dsnbresult}  (right) shows LZ's limits and future sensitivity on the fundamental $\nu_x$ emission parameters, characterized by the energy emitted per $\nu_x$ flavor and the average neutrino energy. The signal model used in the measurement is a simple DSNB model from Ref.~\cite{Suliga:2021hek}, which assumes all supernovae emit the same thermal neutrino spectrum regardless of their NS or BH outcomes. 

\section{Conclusions}
We have performed the first DSNB $\nu_x$ search through the CE$\nu$NS process using data from LZ's first science run. The LZ detector exposure and data selections used in this paper are the same as those from the experiment’s first WIMP search results paper. 
With an exposure of 60~live days using a fiducial mass of 5.5~t, LZ sets an upper limit on the DSNB $\nu_x$ flux that is comparable to the current best limit (at the level of approximately 10$^3$~cm$^{-2}$s$^{-1}$, but using different DSNB models and conventions than those presented in this proceeding and Ref.~\cite{Suliga:2021hek} to set limits) from 1,496 days of SK data using 22.5~kton of fiducial mass~\cite{Lunardini:2008xd}, which demonstrates the power of probing DSNB $\nu_x$ through the CE$\nu$NS process in xenon detectors. 

The LZ limits shown in Fig.~\ref{fig:dsnbresult} do not restrict any existing realistic DSNB model but could be useful in the future. In some astrophysical models, the flux could be larger, for example, if a more massive neutron star or a black hole is formed. In some new-physics models, the mean energy could also be larger, for example, if neutrinos can escape more readily from the core of the proto-neutron star, which would increase the average neutrino energy and hence the detectability. Either or both of these scenarios would improve the prospects for detection in future xenon experiments~\cite{discussion}.

\section*{Acknowledgments}
Q. Xia would like to thank A.M. Suliga and J.F. Beacom for helpful conversations and extensive correspondence on details about DSNB modeling, as well as for providing the code for generating existing limits on the DSNB. The research supporting this work took place in part at the Sanford Underground Research Facility (SURF) in Lead, South Dakota. Funding for this work is supported by the U.S. Department of Energy, Office of Science, Office of High Energy Physics, U.S. National Science Foundation (NSF), the UKRI's Science \& Technology Facilities Council, Portuguese Foundation for Science and Technology (FCT), the Institute for Basic Science, Korea, and the Swiss National Science Foundation (SNSF). This research was supported by the Australian Government through the Australian Research Council Centre of Excellence for Dark Matter Particle Physics. We acknowledge additional support from the UK Science \& Technology Facilities Council (STFC) for PhD studentships and the STFC Boulby Underground Laboratory in the U.K., and the GridPP and IRIS Collaborations. We acknowledge the South Dakota Governor's office, the South Dakota Community Foundation, the South Dakota State University Foundation, and the University of South Dakota Foundation for use of xenon. We also acknowledge the University of Alabama for providing xenon.

\section*{References}

\bibliography{ref}

\end{document}